\documentclass[aps,prb,preprint,superscriptaddress,showpacs]{revtex4-1}
\usepackage[english]{babel}
\usepackage[utf8]{inputenc} 
\usepackage{helvet}         
\usepackage{courier}        
\usepackage{type1cm}        
\usepackage{makeidx}        
\usepackage{graphicx}       
\usepackage[bottom]{footmisc}
\usepackage{amssymb,amsmath}
\usepackage{bm}  
\usepackage{mathrsfs}
\usepackage{subeqnarray} 
\usepackage[pdftex,bookmarks,colorlinks,breaklinks]{hyperref}  
\usepackage{bbold}

\usepackage{color}
\definecolor{dullmagenta}{rgb}{0.4,0,0.4}   
\definecolor{darkblue}{rgb}{0,0,0.4}
\hypersetup{linkcolor=red,citecolor=blue,filecolor=dullmagenta,urlcolor=darkblue} 

\begin{document}

\title{A group action principle for Nambu dynamics of spin degrees of freedom}

\author{Stam Nicolis}
\email{stam.nicolis@lmpt.univ-tours.fr}
\affiliation{CNRS-Institut Denis--Poisson (UMR 7013), Université de Tours, Université d'Orléans, Parc de Grandmont, F-37200, Tours, FRANCE, \email{stam.nicolis@lmpt.univ-tours.fr}}

\author{Pascal Thibaudeau}
\email{pascal.thibaudeau@cea.fr}
\affiliation{CEA DAM/Le Ripault, BP 16, F-37260, Monts, FRANCE}

\author{Thomas Nussle}
\email{thomas.nussle@cea.fr}
\affiliation{CEA DAM/Le Ripault, BP 16, F-37260, Monts, FRANCE}
\affiliation{CNRS-Institut Denis--Poisson (UMR 7013), Université de Tours, Université d'Orléans, Parc de Grandmont, F-37200, Tours, FRANCE, \email{stam.nicolis@lmpt.univ-tours.fr}}

\date{\today}

\begin{abstract}
We describe a formulation of the group action principle, for linear Nambu flows, that explicitly takes into account all the defining properties of Nambu mechanics and  illustrate  its relevance by showing how it can be used to describe the off--shell states and superpositions  thereof that define  the transition amplitudes for the quantization of Larmor precession of a magnetic moment. It highlights the relation between the fluctuations of the longitudinal and  transverse components of the magnetization. 
This formulation has been shown to be consistent with the approach that has been developed in the framework of the non commutative geometry of the 3--torus.
In this way the latter can be used as a consistent discretization of the former.
\end{abstract}


\maketitle

\section{Introduction}\label{intro}
Nambu mechanics is the generalization of Hamiltonian mechanics to phase spaces of arbitrary dimension~\cite{nambu_generalized_1973}.
The reason it is useful to consider such spaces at all is to describe the dynamics of extended objects \cite{bergshoeff_supermembranes_1987} (which was Nambu's original motivation \cite{nambu_hamilton-jacobi_1980}).
It represents, in fact, the generalization of the area preserving diffeomorphisms of Hamiltonian mechanics to the corresponding group(s) of transformations that preserve the volume in spaces of odd dimension, too \cite{guendelman_volume-preserving_1995}.
In this regard it appears much less ``exotic'' and, indeed, has found many applications in classical fluid mechanics, where the study of incompressible flows is the natural context~\cite{nevir_nambu_1993,guha_applications_2002}.

While it became the subject of interest in the effort to understand the dynamics of multiple M2--branes~\cite{ho_nambu_2016}, it was quickly realized that there were conceptual issues that remain to be clarified for the description of their quantum effects.
A workaround that used only generalized Poisson brackets, i.e. a purely Hamiltonian formulation in a non-flat metric, was found to be sufficient for many cases of practical interest \cite{tezuka_poisson_2002}; however a deeper understanding of the properties of the models proposed for M2--branes must solve the problem of the consistent definition of the quantization of the Nambu bracket \cite{awata_quantization_2001,minic_nambu_2002,curtright_classical_2003}, which is a generalization of the Poisson bracket with more than 2 elements.
To this end it is useful to understand the properties of simpler quantum systems, that can be described with the framework of Nambu mechanics. Such examples are provided by magnetic systems\cite{duan_geometric_2001},
where the three components of the magnetization, are, naturally, identified with the canonical variables of Nambu mechanics. 

In this contribution we shall show that a recently proposed quantization scheme~\cite{axenides_nambu_2009} can be applied to describe the quantum dynamics of Larmor precession of a magnetic moment in an external field and can describe the off--shell states in a way that provides insights that are much harder to grasp using the traditional Hamiltonian formalism.

What has been lacking, indeed, is a consistent group action principle, that leads to the definition of consistent unitary, linear, evolution operators that, acting on the space of states, describe consistent superpositions of states and their transition probabilities.
While the proposal was set forth in ref.~\cite{axenides_nambu_2009}, it does require fleshing out for concrete applications, such as the one discussed here.

Therefore, in the following section, we shall review the salient properties of classical linear Nambu flows in the continuum, focusing on the volume preserving diffeomorphisms; we shall, then, show how the consistent quantization of these can be understood in terms of the properties of the non--commutative 3--torus.
We shall then construct the unitary evolution operators on it, implementing a regularization in terms of finite dimensional matrices and show that the size of the matrices has a physical basis.
We conclude with a discussion of further avenues of inquiry, in particular, regarding consistent coupling to baths.

\section{Linear Nambu flows for classical and quantum magnets}\label{LNF}
In 3-dimensional Nambu mechanics, linear Nambu flows are defined by the time evolution equations for each component of the vector of dynamical variables
\begin{equation}
\label{LNFeom}
\frac{dx^I}{dt}=\left\{x^I,H_1,H_2\right\}={\sf M}^{IJ}x^J(t),
\end{equation}
with ${\sf M}$ a constant antisymmetric matrix and the Nambu 3-bracket is defined as
\begin{equation}
\label{bracket1}
\left\{f,g,h\right\}=\varepsilon^{IJK}\partial_I f\partial_J g\partial_K h,
\end{equation}
with $f$, $g$, $h$ are any given functions of ${\bm x}$, $\partial_I\equiv\partial/\partial x^I$ and $\varepsilon^{IJK}$ is the fully anti-symmetric Levi-Civita pseudo-tensor of rank $3$.

The solution of eq.~(\ref{LNFeom}) can be written as
\begin{equation}
\label{LNFeomsol}
\bm{x}(t)=e^{{\sf M}t}\bm{x}(0)\equiv {\sf A}(t)\bm{x}(0)
\end{equation}
Since ${\sf M}$ is traceless, ${\sf A}(t)\equiv e^{{\sf M}t}$, the classical, one--step evolution operator, can be shown to be an element of the group $SL(3,\mathbb{R})$.

Linear systems can be defined by two conserved quantities, $H_1=\bm{a}\cdot\bm{x}$ with $\bm{a}$ a constant vector and $H_2=(1/2)\bm{x}^\mathrm{T}{\sf B}\bm{x}$, with ${\sf B}$ a constant symmetrical matrix. In the framework of Nambu mechanics, these systems define the simplest systems to consider\cite{axenides_nambu_2009} and are formal analogues of the harmonic oscillator from Hamiltonian mechanics.

Such systems are not only toy models, but can  be considered as prototypes for modeling dynamics of magnets,  dominated by the exchange interaction. For instance, if an anti-ferromagnetic material is defined by a magnetic crystalline cell that can be mapped on two sublattices with spins ${\bm s}_1$ and ${\bm s}_2$, then for each spin in its first neighbor cell, without any further interaction, we have the following equations of motion~:
\begin{subeqnarray}
\frac{ds_1^I}{dt}&=&\varepsilon^{IJK}J_{12}s_2^Js_1^K\\
\frac{ds_2^I}{dt}&=&\varepsilon^{IJK}J_{12}s_1^Js_2^K
\end{subeqnarray}
The average ferromagnetic magnetization vector ${\bm M}\equiv\frac{1}{2}\left({\bm s}_1+{\bm s}_2\right)$ and the anti-ferromagnetic Néel vector ${\bm m}\equiv\frac{1}{2}\left({\bm s}_1-{\bm s}_2\right)$ can be defined, and for these vectors we immediately have
\begin{subeqnarray}
  \frac{dM^I}{dt}&=&0\slabel{average-M}\\
  \frac{dm^I}{dt}&=&2J_{12}\varepsilon^{IJK}M^Jm^K\slabel{diff-m}
\end{subeqnarray}
Because ${\bm M}$ is then a constant of motion, eq.(\ref{diff-m}) describes a linear Nambu flow for the antiferromagnetic vector ${\bm m}$, associated with a traceless matrix ${\sf M}$ introduced in eq.(\ref{LNFeom}).

As in Hamiltonian mechanics, this flow is on the phase space of the system and describes, through Liouville's theorem, a flow for the probability density $\rho({\bm x},t)$ therein:
\begin{equation}
\label{Liouvilleclass}
\frac{\partial\rho({\bm x},t)}{\partial t} = \left\{\rho({\bm x},t),H_1,H_2\right\}
\end{equation}

Of particular interest are the moments of this probability density and their evolution in time, since they can be related to observable quantities.

For linear Nambu flows this equation takes the form
\begin{equation}
\label{LiouvilleNambuclass}
\frac{\partial\rho({\bm x},t)}{\partial t}=\varepsilon^{IJK}\omega_JB_{KL}\partial_I\rho({\bm x},t)  x^L=\mathrm{det}\left({\bm\nabla}\rho({\bm x},t),\bm{\omega},{\bm\nabla} H_2\right)
\end{equation}
and describes, also, the classical dynamics in phase space, i.e. the properties of the solutions of the classical equations of motion.

The problem here is that, in Hamiltonian mechanics, it is known that Poisson brackets are the classical limits of commutators \cite{yaffe_large_1982}.
In Nambu mechanics what is the quantum structure that preserves all its properties, whose classical limit would be the Nambu bracket, is not known~\cite{curtright_classical_2003}.

In ref.~\cite{axenides_nambu_2009} the off--shell states and consistent evolution operators for classical and quantum, linear Nambu flows were  constructed.

Let us review the idea of the construction. It is based on introducing an infrared cutoff, by compactifying the phase space on a 3--torus, $\mathbb{T}^3$; and on introducing a short--distance (``ultraviolet'') cutoff, by considering only points with rational coordinates and common denominator, $N$.

In this way the differential equations become linear recurrences on the finite field, $\mathbb{Z}_N$
 \begin{equation}
 \label{LMapsmodN}
 \bm{x}_{n+1}={\sf A}\bm{x}_n\,\mathrm{mod}\,N
 \end{equation}
which can display quite complex, indeed, deterministic chaotic, behavior as $N$ varies.
Already, at the classical level, this means that the system possesses a finite number of states and how these are visited during the evolution is of interest.
If there is periodic behavior, the period, $T(N)$, satisfies ${\sf A}^{T(N)}\equiv I\,\mathrm{mod}\,N$; and the quantum counterpart, $U({\sf A})$ shares this property, by construction, since $U({\sf A}^{T(N)})=[U({\sf A})]^{T(N)}$.
What happens as $N\to\infty$ is a quite delicate issue, that has been investigated in the context of quantum chaos~\cite{faure_scarred_2003}, but there is, still, much to be clarified.

The integer $N$ controls in this way the fluctuations at both ends, infrared and ultraviolet and $2\pi/N$ plays, indeed, the role of Planck's constant for describing the quantum fluctuations~\cite{axenides_nambu_2009}.

The construction of the quantum evolution operator, $U({\sf A})$, on the 3--torus proceeds, in fact, in complete analogy to the Hamiltonian quantization of toroidal phase spaces.
The idea will be to construct a unitary operator, $U({\sf A})$, that realizes a consistent quantization, of the classical evolution operator, ${\sf A}\in SL(3,\mathbb{R})$,  in the sense that it satisfies the correspondence principle--which means that it realizes the metaplectic representation 
--and provides a faithful representation, in the sense that, for any classical evolution operators, ${\sf A}$ and ${\sf B}$, we have the composition rule that
\begin{equation}
\label{Ugroup}
U({\sf A}\circ{\sf B}) = U({\sf A})\circ U({\sf B})
\end{equation}
This property is necessary to ensure that time evolution is well--defined, that it depends only on the endpoints in phase space and not on the parametrization of the path(s).

For a finite-dimensional representation $N$ of the operator ${\sf{A}}$, as sketched in ref.~\cite{axenides_nambu_2009}, the construction of an $N\times N$ matrix $U({\sf A})$ with these properties is realized by showing that it can be mapped exactly to the construction of the corresponding unitary operator of a Hamiltonian system.

This proceeds as follows: first, the linear Nambu flow has the property that $\bm{\omega}$ is an eigenvector of the one--step evolution operator, ${\sf A}$, with eigenvalue 1:
\begin{equation}
\label{fixedevector}
{\sf A}\bm{\omega}=\bm{\omega}\Leftrightarrow \bm{\omega}^T=\bm{\omega}^T{\sf A}
\end{equation}
The convention of multiplication from the right was used (i.e for the dual vectors) in ref.~\cite{axenides_nambu_2009}; in the present contribution we use the, perhaps, more familiar convention of multiplication from the left.
Because one deals with finite dimensional ordinary and possibly complex valued vectors, it does not really matter, up to transposition and conjugate.
The property (\ref{fixedevector}) is also true for the any collinear vector $\lambda\bm{\omega}$ that form an infinite collection of fixed vector with $1$ as eigenvalue.

The property that the vector of the linear Hamiltonian is left invariant by the flow implies, in turn, for eq.~(\ref{LMapsmodN}), that
\begin{equation}
\label{Hamevol}
\left[\bm{x}\times\bm{\omega}\right]_{n+1} = {\sf A}[\bm{x}\times\bm{\omega}]_n
\end{equation}
for any time step $n$.

This expression can now be shown to be equivalent to
\begin{equation}
\label{Hamevol1}
\left[\bm{x}\times\bm{\omega}\right]_{n+1} = \left(\begin{array}{cc} \widetilde{{\sf A}} & \bf{0}\\
\widetilde{\bm{\omega}} & 0\end{array}\right)
[\bm{x}\times\bm{\omega}]_n
\end{equation}
which provides a definition of the $2\times 2$ evolution operator $\widetilde{\sf A}$.
It has been shown in ref.~\cite{thibaudeau_thermostatting_2012} that one can construct a basis using the initial magnetization state $\bm{x}(0)$, the precession vector $\bm{\omega}$ and their vector product $\bm{x}(0)\times \bm{\omega}$, and it is possible to decompose the time evolution of the solution on this basis as
\begin{equation}
	\bm{x}(t)=A(t)\bm{x}(0)+B(t)\frac{\bm{\omega}}{\|\bm{\omega}\|}+C(t)\bm{x}(0)\times \frac{\bm{\omega}}{\|\bm{\omega}\|}
\end{equation}
As long as the precession vector $\bm{\omega}$ is constant, one can always choose a reference frame such that said vector is aligned with the $\bm{z}$-axis. If we now consider the time evolution of $\bm{x}(t)\times \bm{\omega} $ one can see that this vector remains in the $(\bm{x},\bm{y})$ plane. As such the last component remains null over time. Hence one can restrain eq.~(\ref{Hamevol}) to eq.~(\ref{Hamevol1})

This can be shown to be symplectic, therefore the corresponding quantum evolution operator, $U(\widetilde{\sf A})$,  can be constructed by known techniques.
It is this operator that we shall define as the unitary evolution operator of the quantum Nambu evolution.
In subsequent sections we shall show that our construction passes a non--trivial test, by checking that it provides results that are consistent with those obtained by the canonical quantization of the Larmor precession.

The off--shell states are, therefore, those that are defined by the action of operators mod $N$, whereas the on--shell states are those that do not require the mod $N$ operation. However, quantum effects are, also, described by superpositions of pure states. We shall show the relevance of such superpositions in the following section. 

\section{Computing transition probabilities à la Nambu}\label{ProbaLNF}
In this section we shall show how to use the unitary evolution operator, $U({\sf A})$ to compute transition probabilities for Larmor precession, from any initial to any final state of the magnetic moment.

Our starting point is the identification of the Larmor precession equation as  a linear Nambu flow, following the notation of ref. \cite{axenides_nambu_2009}
\begin{equation}
	\frac{dx^I}{dt}=\epsilon^{IJK}a_JB_{KL}x^L\Leftrightarrow
	\frac{ds^I}{dt}=\epsilon^{IJK}\omega_J s_K\equiv {\sf M}^{IK}s_K
\end{equation}
Where $H_1\equiv \bm{a}\cdot{\bm{x}}=\bm{\omega}\cdot\bm{s}$ and $H_2\equiv (1/2)(\bm{x},{\sf B}\bm{x})$, with ${\sf B}=\mathbb{1}$.

For Larmor precession around an external field described by $\bm{\omega}$ we, thus,  have
\begin{equation}
	{\sf M}={\bm \omega}\times.=\left(\begin{array}{ccc}
	0 & -\omega_3  &\omega_2  \\
	\omega_3& 0 &-\omega_1  \\
	-\omega_2&\omega_1  &0
	\end{array} \right)
\end{equation}
If Larmor precession happens around a fixed vector $\bm{\omega}$, we can always choose our reference frame such that only the component along the $z$-axis is non-zero $\bm{\omega}=(0,0,\omega_3)$.

Its exponential, ${\sf A}=\exp({\sf M})$ is the one--step evolution operator.
This acts on a finite set of states, labeled by the integers  mod $N$~\cite{axenides_nambu_2009}, which means that the matrix ${\sf A}\in SL(3,\mathbb{Z}_N)$, which, also, has integer entries, mod $N$, has the form
\begin{equation}
A=\left(\begin{array}{ccc}
	a & b  & 0 \\
	-b & a & 0  \\
	0 & 0  &1
	\end{array} \right)
\end{equation}
where $a$ and $b$ are integers mod $N$, which satisfy $a^2+b^2\equiv\,1\  \mathrm{mod}\, N$. We may, hence, work out the form of the ``reduced" evolution operator $\tilde{A}\in SL(2,\mathbb{Z}_N)$
\begin{equation}
\label{tildeA}
\tilde{A}=\left(\begin{array}{ccc}
	a & b   \\
	-b & a
	\end{array} \right)
\end{equation}
If the three--component states are labeled by the vector $\bm{s}$, then the ``reduced''  states are labeled by the  vector $\bm{\tilde{s}}$
\begin{equation}
\bm{\tilde{s}}=(\omega_3 s_1-\omega_1 s_3,\omega_3 s_2-\omega_2 s_3)
\end{equation}
Now by choosing the unit of time appropriately, we may set $\omega_3\equiv\,1\,\mathrm{mod}\,N$.

 The previous expression thus becomes
\begin{equation}
\bm{\tilde{s}}=(s_1,s_2)\,\mathrm{mod}\,N
\end{equation}
This means that all the interesting dynamics happens on a plane, orthogonal to the magnetization vector $\bm{\omega}$ and as such, once we choose an initial state with fixed value for $s_3$ we have to satisfy $s_1^2+s_2^2=(1-s_3^2)\ \mathrm{mod}\, N$. 

What is interesting in this expression is that, if $1-s_3^2$ is a quadratic residue mod $N$, this expression can be reduced to $\sigma_1^2 + \sigma_2^2\equiv\,1\,\mathrm{mod}\,N$. If not, we must work in the quadratic extension of the number field. In this way the transverse fluctuations, described by $s_1$ and $s_2$ are related to the longitudinal fluctuations, described by $s_3$. 
  We now have a way to count all the states which are accessible from any  one initial state and as such we can label them.

All that remains to be computed, therefore,  is the quantum evolution operator $U(\tilde{\sf A})$, whose classical limit would be $\widetilde{{\sf A}}$.

According to reference~\cite{axenides_nambu_2009}, 

\begin{equation}
\label{UofALarmor}
\left[U(\widetilde{\sf A})\right]_{k,l}=\frac{(2b|N)}{\sqrt{N}}\Omega_N^{\frac{ak^2-2kl+dl^2}{2b}}
\end{equation}
where $\Omega_N=e^{2\pi\mathrm{i}/N}$ is the $N-$th root of unity, and $(2b|N)$ is the Jacobi symbol, for $2b$ and $N$, equal to 1 if $2b$ is a quadratic residue mod $N$, $-1$ if not and 0 if $2b\equiv 0\,\mathrm{mod}\,N$. 

 While these expressions were originally derived for $N$ prime, it has been shown  that the matrices factorize over the prime factorization of $N$~\cite{athanasiu_fast_1998}.

Both the time evolution of the quantum states $|\bm{\tilde{s}}\rangle$ and the transition probabilities between them are given by the evolution operator as
\begin{eqnarray}
	& |\bm{\tilde{s}}\rangle_n\equiv U(\widetilde{\sf A}^n)|\bm{\tilde{s}}\rangle_0 \\
	& P_n(\bm{\tilde{s'}},\bm{\tilde{s}})=|\langle\bm{\tilde{s'}}|U(\widetilde{\sf A}^n)|\bm{\tilde{s}}\rangle|^2
\end{eqnarray}

Let us  illustrate this abstract  framework with a specific  example by taking  $N=5$, $\omega_3=1$, and the initial magnetization state to be normal to the external field described by $\bm{\omega}$ (i.e in the $(\bm{x},\bm{y})$ plane, so  that $s_3\equiv 0$). This means all the accessible states are those for which
\begin{equation}
	s_1^2+s_2^2\equiv\,1\, \mathrm{mod}\ 5
\end{equation}
To count and label them we have
\begin{eqnarray*}
	|1\rangle=(1,0) \quad
	|2\rangle=(0,1) \quad
	|3\rangle=(4,0) \quad
	|4\rangle=(0,4)
\end{eqnarray*}
We note that these are, also, ``classical'' states. Quantum effects are described by their superpositions, that don't have a classical analog. 

Furthermore, the only three, non--trivial, evolution operators, $\widetilde{\sf A}$, satisfying the constraint  $a^2 + b^2\equiv\,1\, \mathrm{mod}\ 5$ are
\begin{eqnarray*}
\tilde{A}_1=\left(\begin{array}{ccc}
0 & 1   \\
4 & 0
\end{array} \right)
\text{, }
\tilde{A}_2=\left(\begin{array}{ccc}
	0 & 4   \\
	1 & 0
\end{array} \right)
\text{and }
\tilde{A}_3=\left(\begin{array}{ccc}
	4 & 0   \\
	0 & 4
\end{array} \right)=(\tilde{A}_1)^2
\end{eqnarray*}
These matrices describe rotations by $\pm 90^\mathrm{o}$ in phase space--the Fourier transform.
This means that the quantum evolution operator, $U(\widetilde{\sf A})$ is, in fact, the Discrete Fourier Transform, over five states. This, apparently, is one more state than necessary, since $\widetilde{\sf A}^4=\mathbb{1}_{2\times 2}\Leftrightarrow U(\widetilde{\sf A})^4=U(\widetilde{\sf A}^4)=U(\mathbb{1}_{2\times 2})=\mathbb{1}_{5\times 5}$. 
This, of course, means that the states are degenerate--their degeneracies were studied in detail by Mehta~\cite{mehta_eigenvalues_1987}. 

However, any state $|\bm{\tilde{s}}\rangle$, can be expanded in the basis of the eigenstates of $U(\widetilde{\sf A})$
\begin{equation}
\label{eigenexpansion}
|\bm{\tilde{s}}\rangle = \sum_{k=0}^{N-1}\,c_k|\psi_k\rangle
\end{equation}
These states are superpositions of the states of definite magnetization. 
This, however, means that there are only $N-1$ independent, relative  phases, since the evolution operator is unitary. Therefore there are only $N-1=4$, in the case at hand, ``non--trivial'' states.  So, let us label the additional state as $|0\rangle$. To define a more convenient way of dealing with the superposition of states, we will use the following notation
\begin{equation}
	|\alpha,\beta,\gamma,\delta,\epsilon\rangle\equiv \alpha|0\rangle+\beta|1\rangle+\gamma|2\rangle+\delta|3\rangle+\epsilon|4\rangle
\end{equation}
To compute transition probabilities depending on time (the integer $n$ playing the role of a discrete time evolution here, hence a ``kicked"-precession), if  we start with an initial state $|0,1,0,0,0\rangle\equiv|1\rangle$, in the basis of the position operator,  after one time-step, the next state is given by

\begin{equation}
	|\bm{\tilde{s}}\rangle_1=U(\tilde{A})|\bm{\tilde{s}}\rangle_0
\end{equation}

The evolution operator $U(\widetilde{{\sf A}}_1)$, describing the Discrete Fourier Transform, 
applied to the initial state $|1\rangle$
\begin{equation}
\label{next_state}
	U(\tilde{A}_1)|1\rangle =\frac{1}{\sqrt{5}}\left[-1|0\rangle+e^{\frac{3\imath\pi}{5}}|1\rangle+ e^{\frac{4\imath\pi}{5}}|2\rangle] - e^{\frac{1\imath\pi}{5}}|3\rangle-e^{\frac{2\imath\pi}{5}}|4\rangle\right]
\end{equation}
where we have highlighted the relative phases, of the other pure states wrt  the state $|0\rangle$.
 
\begin{figure}[htp]
	\centering
	\resizebox{0.7\columnwidth}{!}{\includegraphics{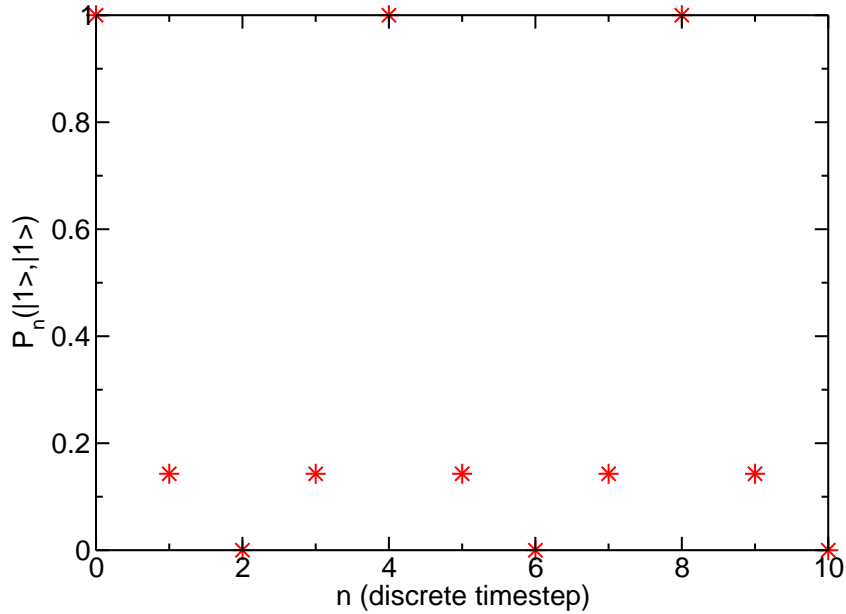}}
	\caption{
	Transition probability  $P_n(|1\rangle,|1\rangle)=|\langle 1| U(\tilde{A}_1^n)|1\rangle|^2$ as function of the time-step $0\leq n\leq 10$ highlighting the periodicity. \label{Fig1}}
\end{figure}

The time evolution of the transition probability between the same initial and final pure state, say $|1\rangle$, as a function of the discrete time-step $n$
\begin{equation}
	P_n(|1\rangle,|1\rangle)=|\langle1|U(\tilde{A}_1)^n|1\rangle|^2
\end{equation}
computes the Nambu path integral and 
should display the appropriate periodicity, viz. 
\begin{equation*}
	U(\tilde{A}_1)^{T(N)}=\mathbb{1}_{5\times5}
\end{equation*}
Results are displayed in Figure~\ref{Fig1}.

\section{Conclusion and outlook}\label{outlook}
In this contribution we have proposed a group action principle for linear Nambu flows, that is consistent with the properties of classical Nambu mechanics, as well as the correspondence principle of quantum mechanics and can, thus, be considered as a consistent quantization of linear Nambu flows. Our formalism provides an explicit prescription for the space of states, both ``on--shell'' and ``off--shell'' and linear superpositions, that are the hallmark of non--classical behavior. Such flows are relevant for describing the Larmor precession of the magnetization of nanamagnets  and, thus, their quantization is relevant for describing its  quantum fluctuations. We have applied our framework to the calculation of transition probabilities and computing the time evolution of a simple model for quantum Larmor precession mod 5, that is relevant for a spin 2 nanomagnet.

The semi--classical limit can be obtained, as $N$, therefore the number of spin states becomes large, as might be expected and is quite subtle.

When Gilbert damping is taken into account, the equations of motion become non--linear, but can, still be solved; their solutions can  be interpreted as describing instantons.
Interestingly the damped linear Nambu flow admits a continuous evolution solution \cite{thibaudeau_thermostatting_2012} that it would be of practical interest to set  in the presented framework.



\bibliographystyle{apsrev4-1}

\end{document}